# Revealing the Brønsted-Evans-Polanyi Relation in Halide-Activated Fast MoS$_2$ Growth Towards Millimeter-Sized 2D Crystals

Qingqing Ji[1,10], Cong Su[2,3,4,5,*], Nannan Mao[6], Xuezeng Tian[7,8], Juan-Carlos Idrobo[9], Jianwei Miao[7,8], William A. Tisdale[6], Alex Zettl[3,4,5], Ju Li[2] and Jing Kong[1,*]

[1] Department of Electrical Engineering and Computer Science, Massachusetts Institute of Technology, Cambridge, Massachusetts 02139, United States

[2] Department of Nuclear Science and Engineering, Massachusetts Institute of Technology, Cambridge, Massachusetts 02139, United States

[3] Kavli Energy NanoSciences Institute at the University of California, Berkeley, CA 94720, United States

[4] Department of Physics, University of California, Berkeley, CA 94720, United States

[5] Materials Sciences Division, Lawrence Berkeley National Laboratory, Berkeley, CA 94720, United States

[6] Department of Chemical Engineering, Massachusetts Institute of Technology, Cambridge, Massachusetts 02139, United States

[7] Department of Physics and Astronomy, University of California, Los Angeles, CA 90095, United States

[8] California NanoSystems Institute, University of California, Los Angeles, CA 90095, United State

[9] Center for Nanophase Materials Sciences, Oak Ridge National Laboratory, Oak Ridge, TN 37831, United States

[10] Present address: School of Physical Science and Technology, ShanghaiTech University, Shanghai 201210, China

*Corresponding author: csu@berkeley.edu; jingkong@mit.edu

# ABSTRACT

Achieving large-size two-dimensional (2D) crystals is key to fully exploiting their remarkable functionalities and application potentials. Chemical vapor deposition (CVD) growth of 2D semiconductors such as monolayer $MoS_2$ has been reported to be activated by halide salts, yet clear identification of the underlying mechanism remains elusive. Here we provide unambiguous experimental evidence showing that the $MoS_2$ growth dynamics are halogen-dependent through the Brønsted-Evans-Polanyi relation, based on which we build a growth model by considering $MoS_2$ edge passivation by halogens, and theoretically reproduces the trend of our experimental observations. These mechanistic understandings enable us to further optimize the fast growth of $MoS_2$ and reach record-large domain sizes that should facilitate practical applications.

Two-dimensional (2D) semiconductors such as monolayer $MoS_2$ are essential building blocks for next-generation ultrathin flexible and low-power electronics[1,2]. As a premise for this, their batch production requires large domain size[3,4], large-area continuity, and thickness uniformity[5,6]. These are difficult tasks in particular for synthetic $MoS_2$ and other transition metal dichalcogenides (TMDs), considering the less controllable mass flux from solid metal precursors[7] in chemical vapor deposition (CVD), as compared to the case of graphene growth with gaseous hydrocarbons[8]. One recent advance to mitigate this is the use of alkali metal halide salts (*e.g.*, NaCl, KI, and NaBr[9–11]), which, in conjunction with transition metal or metal oxide powders, could increase the mass flux of metal precursors and accelerate the 2D growth of TMDs. However, the detailed mechanism remains unclear: both the alkali metal ions[12] and the halogens[10,13] have been proposed to be the key for the promoted growth. Experimental evidences are hence necessary to address this ambiguity and could provide insights to further engineer the growth for desired morphological properties. This is challenging in particular for volatile halogens that barely exist in post-growth samples, thus precluding most ex-situ characterization techniques.

While both the alkali metal-assisted and halogen-assisted mechanisms are plausible, we here provide unambiguous experimental evidence that halogens are closely related to $MoS_2$ growth dynamics. To achieve this, post-growth Arrhenius analysis is implemented without any involvement of *in-situ* characterizations. We find that within the same reaction family, the halide-assisted growths conform to Brønsted-Evans-Polanyi (BEP) relation, where their reaction barriers are linearly correlated to the Mo-X (X = I, Br, Cl, F, and O) bond dissociation energies ($E_b$), suggesting the substitution of Mo-X bonds by the Mo-S bonds to be the rate-limiting step for the CVD growth. Based on this, we propose a theoretical growth model that not only reproduces the BEP relation but also explains the sulfur concentration-dependent growth dynamics observed in our experiments. By harnessing the synergistic effect of the KI promoter and the sulfur supply, we can reproducibly synthesize near millimeter-sized 2D $MoS_2$ crystals dispersed over the entire $SiO_2$/Si substrates in a short growth time. These results not only shed lights on the detailed mechanism of TMD growth activated by the halide salts, but also guide the designer growth towards larger domain sizes that should enable practical applications.

Fig. 1a schematically illustrates our method to produce large-size atomically thin $MoS_2$ crystals. In brief, $MoO_3$ dissolved in ammonia is spin-coated on the $SiO_2$/Si substrates[14] and loaded subsequently into a tube furnace for high-temperature annealing under a sulfur atmosphere (see Supplementary Fig. 1 for more experimental details). The $MoO_3$/ammonia solution provides an advantage that soluble salts such as KI can be incorporated into the solution and mix uniformly with the Mo source in the spin-coated film. Under optimized sulfur heating temperature ($T_S$ = 165 °C), large size 2D $MoS_2$ crystals are synthesized and distributed over the entire substrate, directly visualizable even with naked eyes (Fig. 1b). The flake sizes follow a Gaussian distribution centered at ~0.5 mm, which is among the largest synthetic 2D $MoS_2$ domains reported[4,15].

Fig. 1c is an optical image of a typical $MoS_2$ crystal produced using the above method. The bulged triangle has a domain size of ~0.62 mm measured vertex-to-vertex, with atomic force microscopy confirming its monolayer thickness of ~1 nm (inset of Fig. 1c). Corresponding Raman and photoluminescence mapping images of this 2D crystal are provided in Supplementary Fig. 2, demonstrating its microscopic uniformity as a monolayer semiconductor. These flakes are proved to be mostly single crystals through second harmonic generation (SHG) imaging. As shown in Fig. 1d, SHG mapping suggests the absence of any grain boundaries with uniform intensity over 1/3 of the flake region[16]. Polarized SHG patterns under parallel configuration (Fig. 1e) are subsequently used to further verify the identical crystal orientation[17] for the remaining flake area (locations 1-5). The single crystallinity is found to extend over the entire flake until another small domain is encountered in the lower left part (location 6). Additionally, SHG imaging of an entire 2D MoS$_2$ flake (~0.4 mm large) by stitching several mapping images is presented in Supplementary Fig. 3 and directly verifies its single crystallinity.

We further demonstrate experimentally that the domain size enhancement here achieved is due to the lowering of the reaction barrier in the presence of KI promoter, by comparing the MoS$_2$ growth results with and without KI in the spin-coated films. Fig. 2a presents domain size statistics for the two cases under varied growth temperature ($T_{Mo}$). Detailed growth results can be found in Supplementary Figs. 4 & 5. The effect of KI incorporation is evident, with consistently much larger crystal size than for specimens grown without KI (insets of Fig. 2a). Since all the other growth parameters in these experiments are kept identical except $T_{Mo}$, the average domain size $D$, being $T_{Mo}$-dependent and proportional to the crystal growth rate, can be plotted versus $1000/T_{Mo}$ for direct Arrhenius fitting (Fig. 2b):

$$\log(D / D_0) = C - (E_a^{exp} / 1000 k_B \ln 10)(1000 / T_{Mo}), \tag{1}$$

where $D_0 = 1$ μm is for normalization, $C$ is a constant, $E_a^{exp}$ is the experimentally derived reaction barrier, and $k_B$ is Boltzmann's constant. We find that the $E_a^{exp}$ at $T_S = 180$ °C decreases significantly from 2.57 eV to 1.80 eV after KI incorporation. This indicates that the iodide salt modifies the detailed reaction pathway of the CVD process, in addition to increasing the vapor pressure of the Mo-containing species. With the presence of KI and $T_S$ lowering from 180 °C to 170 °C, the $E_a^{exp}$ further reduces to 0.87 eV along with the domain sizes increasing to more than 0.2 mm (Fig. 2b; see Supplementary Fig. 6 for the domain size statistics at $T_S = 170$ °C). Optimizing $T_S$ at 165 °C results in even larger 2D MoS$_2$ crystals (Fig. 1c), but the sulfurization reaction cuts off at $T_S \leq 160$ °C, producing only unsulfurized species[18] (Supplementary Fig. 7). We hence conclude that the MoS$_2$ growth dynamics are sensitive to the presence of KI and the sulfur concentration at $T_S > 160$ °C.

By far, it has been challenging to experimentally differentiate the contributions of cations and anions in the halide salts to the activated MoS$_2$ growth. Following the strategy presented above, we have been able to extract the CVD reaction barriers incorporating KBr, KCl, and KF promoters (Supplementary Fig. 8). All of these $E_a^{exp}$ values are found to correlate linearly with the Mo-X bonding energies, $E_b$(Mo-X)[19] (Fig. 2c), which indicates that halogens play a significant role in tuning the MoS$_2$ growth dynamics. The linear correlation can be understood in the context of the BEP principle[20] that predicts, for chemical reactions of the same class, $E_a = E_0 + \alpha \Delta H$, where $\Delta H$ is the enthalpy of reaction, and $E_0$ and $\alpha$ are linear fitting parameters. In our case, the replacement of $\Delta H$ with $E_b$(Mo-X) holds only if we consider MoX$_y$ + 2S → MoS$_2$ + $y$X, where $\Delta H = y E_b(\text{Mo-X}) - 2 E_b(\text{Mo-S})$, as the relevant reaction. Such a BEP relation is frequently observed in surface-catalyzed chemical processes[21,22] and radical reactions dominated by bond substitution[23], in analogue to which the above CVD reaction can also be inferred as bond-substitution dominated. The process concerning this analysis is shown in the inset of Fig. 2c, schematically illustrating the substitution of Mo-X bonds by Mo-S bonds during MoS$_2$ growth.

Based on the above understandings from experiments, we build a theoretical growth model termed *edge passivation-substitution* (EPS), to further rationalize the reaction barrier tuning by halide salts. From both selected-area electron diffraction[24] (Supplementary Fig. 9) and atomic-resolution scanning transmission electron microscopy (STEM) (Fig. 3a), the edge of the MoS$_2$ crystals produced by halide-assisted growth has been identified to be universally along the crystallographic zigzag-Mo orientation, instead of the zigzag-S orientation. Knowing the edge type is useful to narrow down the possible range of the chemical potential of sulfur ($\mu_S$): lower (higher) $\mu_S$ favors zigzag-Mo (zigzag-S) edge type formation. We note that the edge type here indicates the crystal *orientation* rather than the detailed *atomic structure* of the edge (e.g. "zigzag-Mo orientation" corresponds to both zigzag-Mo and antenna-S structures at the edge, the latter being the sulfur-passivated zigzag-Mo edge).

Fig. 3b is an illustration of the EPS model that exhibits the alternate attachment of S and MoX$_y$ clusters on a zigzag-Mo edge to imitate the MoS$_2$ growth process[25]. The growth without halide promoters is plotted as a reference, where the zigzag-Mo edges are passivated by oxygen atoms (lower part of Fig. 3b), as confirmed in the BEP relation (Fig. 2c) that the reaction barrier in this case correlates with $E_b(\text{Mo=O})$. The thermodynamically allowed edge configurations passivated by halogens and oxygen are determined by comparing their incremental chemical potentials on the zigzag-Mo edge to that in the bulk phases, $\mu^{\text{bulk}}$, which represent the upper limit of the chemical potentials of corresponding elements (Fig. 3c). If the energy gain of attaching an atom to the MoS$_2$ edge is lower than its $|\mu^{\text{bulk}}|$, the edge structure with that additional atom is thermodynamically forbidden. Since the halogen/oxygen atoms are not consumed by the growth reaction, they can be rationally assumed to saturate the edge to its maximum extent, ignoring the kinetic effects. Coordination numbers can thus be derived with the detailed edge structures shown in Fig. 3d.

In our EPS model, the reaction barrier ($E_a^{\text{cal}}$) of zigzag-Mo edge growth is approximated as the formation energy ($E_f$) difference between the sulfur-bonded intermediate state (antenna-S edge) and the halogen/oxygen-passivated zigzag-Mo edge. Likewise, in the case of zigzag-S edge growth, $E_a^{\text{cal}}$ is the difference of $E_f$ between the halogen/oxygen-passivated Mo-bonded intermediate state (antenna-Mo edge) and zigzag-S edge. More rigorously, we define

$$E_a^{\text{cal}} = \max\left\{ \left| E_{\text{f,zigzag-Mo}} - E_{\text{f,antenna-S}} \right|, \left| E_{\text{f,zigzag-S}} - E_{\text{f,antenna-Mo}} \right| \right\}, \tag{2}$$

which takes into account the competitive formation of zigzag-Mo and zigzag-S edges governed by Wulff construction theory[26], since the edge growth with smaller activation energy diminishes quickly and the final growth is determined by the slower edge formation process (Fig. 4a). Detailed discussion of the growth model and the density functional theory calculations can be found in Supplementary Fig. 10 and Tables 1-3.

Fig. 4b exemplifies the calculated results for the growth with iodide promoter (calculations of the growth with passivation of oxygen and other halogens are presented in Supplementary Fig. 11). The overall $\mu_S$-dependent $E_a^{cal}(X=I)$ according to Eq. (2) is highlighted in bold in the right panel, exhibiting a V-shape curve minimized at $\mu_S$ = -4.6 eV. Adopting $\mu_S$ = -4.9 eV, which is on the left branches of all the five $E_a^{cal}(X) - \mu_S$ curves that associates with zigzag-Mo edge formation, as confirmed by the STEM image above, we theoretically reproduce the BEP relation (Fig. 4c) and verify the halogen-dependent $MoS_2$ growth dynamics. Notably, the calculated BEP slope is higher than the experimental counterpart, which possibly associates with the fact that, we do not consider the synergistic Mo-X and Mo-S bond dissociation/formation process in the EPS model for simplicity (see Supplementary information for more detailed discussion). Regarding the linearly correlated $E_a$ and $E_b$(Mo-X), the qualitative consistency between experimental and theoretical results justifies our hypothetical EPS model, hence providing useful insights on further engineering the growth of $MoS_2$ and other TMDs.

Our results suggest that edge passivation by oxygen or other additives, being overlooked in most previous studies, should be considered to better understand the $MoS_2$ growth dynamics. Additionally, while the role of alkali metal ions in the salts is not the primary focus of this work, we anticipate that our growth model could be further extended to incorporate these metal ions, by for example placing them on the zigzag-S and antenna-S edges to modify the corresponding formation energies and the overall reaction barriers.

Comparing with previous efforts on large-domain $MoS_2$ growth[3,4,6,12,15,27–30], our method significantly improves the achievable domain size and shortens the required growth time (Fig. 5). It is worth noting that, while crystal growth does not precisely start (end) at the beginning (ending) of $T_{Mo}$, the time span of $T_{Mo}$ is a good approximation of the growth time and a fair gauge for comparison across different methods. As a result, we enhance the growth velocity of $MoS_2$ by 3-5 folds from the previous record[29]. Reaction dynamics engineering mediated by the halide salts has been proved to be the key to achieve this, which differs from other works that focus on improving the nucleation dynamics (e.g., introducing oxygen for nuclei etching[3,12]) or diffusion dynamics (e.g., using molten glass substrates to facilitate precursor diffusion[12,29]). Using a combination of these growth techniques, further improvement of the $MoS_2$ domain size up to millimeter or even larger scales can be rationally anticipated, which represents a long-term pursuit among the 2D semiconductors community.

In summary, we have explored large-domain $MoS_2$ growth activated by halide salts, and provided direct evidence that halogens associate closely with the reaction dynamics through the BEP relation. This mechanistic understanding has enabled us to establish a theoretical growth model that rationalizes all the experimental observations, as well as to guide the designer CVD growth towards millimeter-sized 2D $MoS_2$ crystals with shortened growth time. We postulate that the uncovered activation mechanism is broadly applicable to the growth of diverse 2D metal chalcogenides, and will be of great utility to further integrated functionalities.

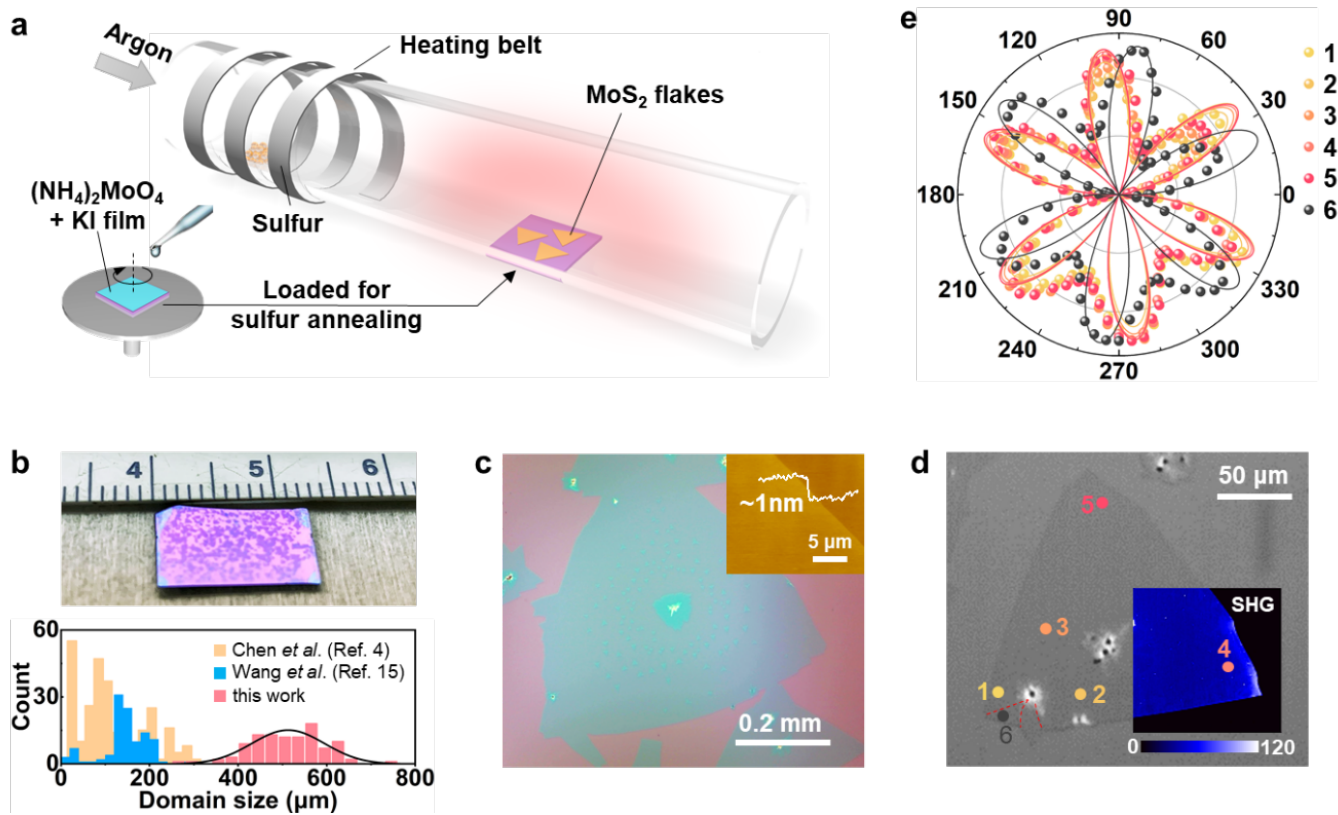

**Fig. 1: Halide-enabled growth of large 2D MoS₂ crystals. a,** Schematic illustration of the growth process. **b,** Photograph of near millimeter-sized 2D MoS₂ crystals over the entire SiO₂/Si substrate (upper panel) and the corresponding domain size distribution (lower panel). The domain size statistics are compared with state-of-the-art achievements on large-domain MoS₂ growth[4,15]. **c,** Optical image of a typical 2D MoS₂ crystal with inset AFM height image on the flake edge showing the monolayer thickness. **d,** SHG mapping of a 2D MoS₂ crystal overlaid on the optical image. Dashed red lines mark the grain boundaries. **e,** Plot of the SHG patterns under parallel configuration on the six locations marked in (**d**). Solid lines are fitted curves.

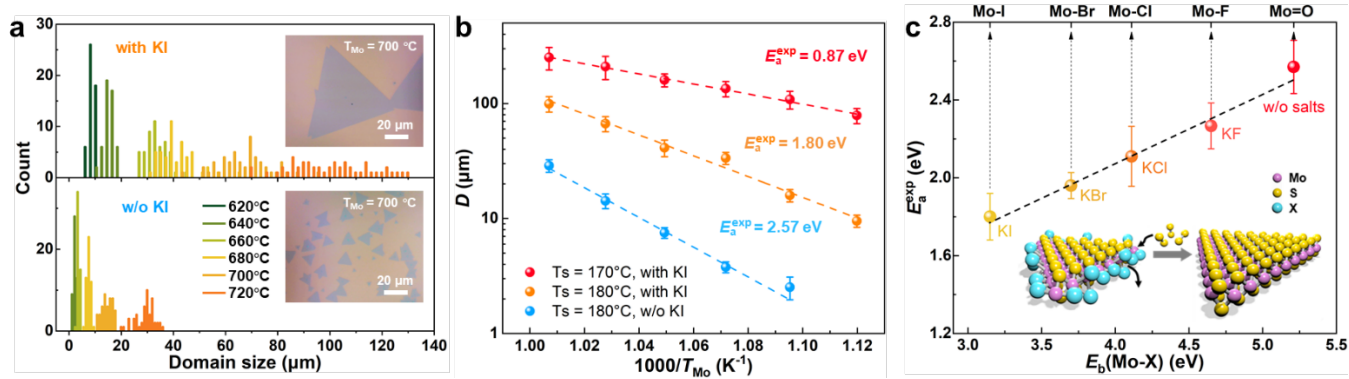

**Fig. 2: Halide-dependent reaction barriers. a,** Statistics of MoS$_2$ domain size under varied growth temperature ($T_{Mo}$), for the growth with and without KI promoters (upper and lower panels, respectively). Insets are the corresponding optical images of MoS$_2$ crystals grown at $T_{Mo}$ = 700 °C for direct comparison. Sulfur heating temperature ($T_S$) was kept at 180 °C. **b,** Arrhenius Plots of average domain size ($D$) versus 1000/$T_{Mo}$ under varied growth conditions. **c,** Plot of the experimentally extracted reaction barriers ($E_a^{exp}$) versus Mo-X bond dissociation energies, $E_b$(Mo-X) (X = I, Br, Cl, F, and O), for the MoS$_2$ growth assisted by various potassium halides. Inset: schematic model of the bond substitution process that dominates MoS$_2$ growth.

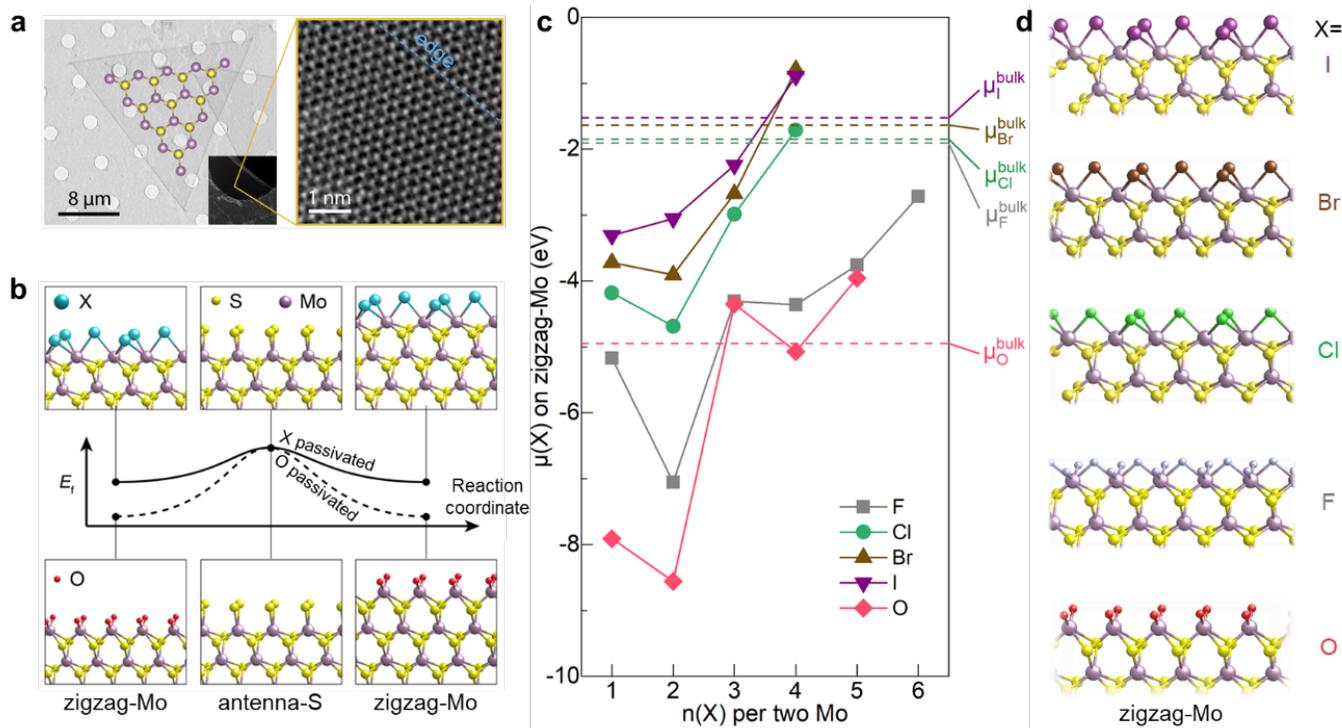

**Fig. 3: Edge passivation-substitution (EPS) model of the halide-assisted MoS₂ growth. a,** TEM and STEM images of the grown MoS₂ flakes. The STEM image is taken close to an edge with the dash line marking the edge orientation. **b,** Energy landscapes and intermediate edge structures of MoS₂ growth with (upper) and without (lower) halide promoters. **c,** Plot of the calculated incremental chemical potentials of halogens and oxygen versus their coordination numbers on the edge. $n$(X) per two Mo is used to accommodate bridge bonding structures. Dash lines mark the corresponding chemical potentials in the bulk phases. **d,** Thermodynamically allowed zigzag-Mo edge structures passivated by maximum halogen and oxygen atoms.

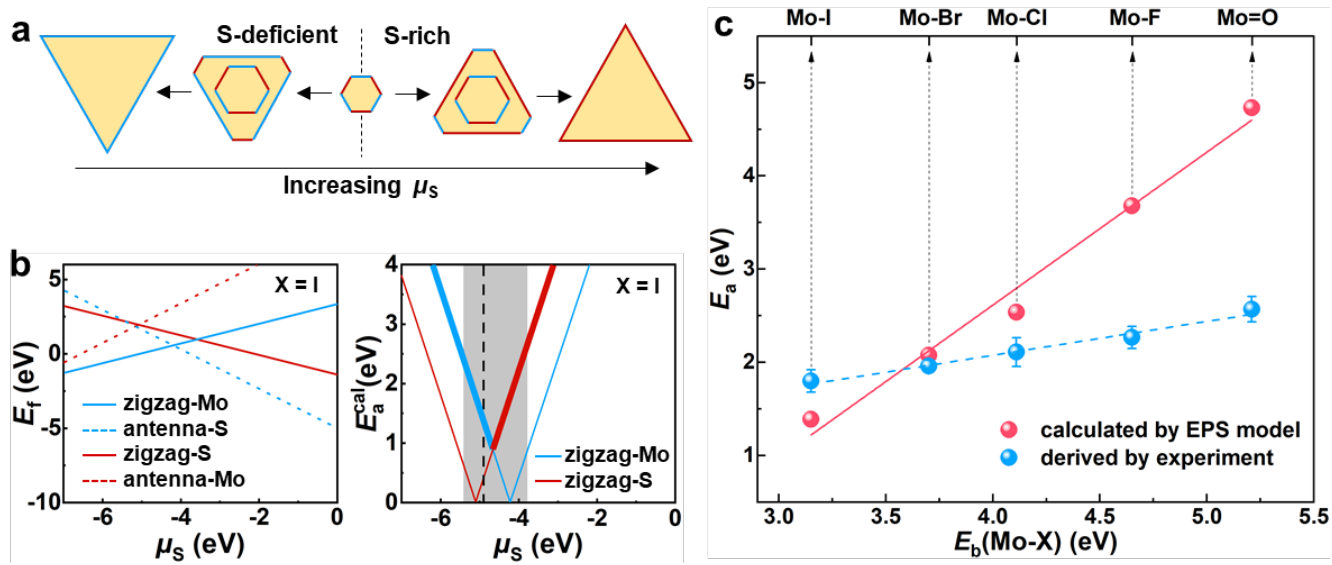

**Fig. 4: Theoretically derived reaction barrier and BEP relation using the EPS model. a,** Schematic shape evolution of $MoS_2$ crystals under S-deficient (Mo-rich) and S-rich conditions. Blue and red lines represent zigzag-Mo and zigzag-S terminated edges, respectively. **b,** Edge formation energy ($E_f$) and derived reaction barrier ($E_a^{cal}$) as a function of the chemical potential of sulfur ($\mu_S$) in the case of iodine passivation. In the right panel, the shaded region indicates accessible $\mu_S$, and the vertical dash line marks where $\mu_S$ = -4.9 eV. **c,** Halide-dependent $E_a^{cal}$ (in red) versus $E_b$(Mo-X) calculated with the above growth model. $\mu_S$ = -4.9 eV is adopted to derive the $E_a^{cal}$. The experimental data (in blue) from Fig. 2c are also provided here for direct comparison.

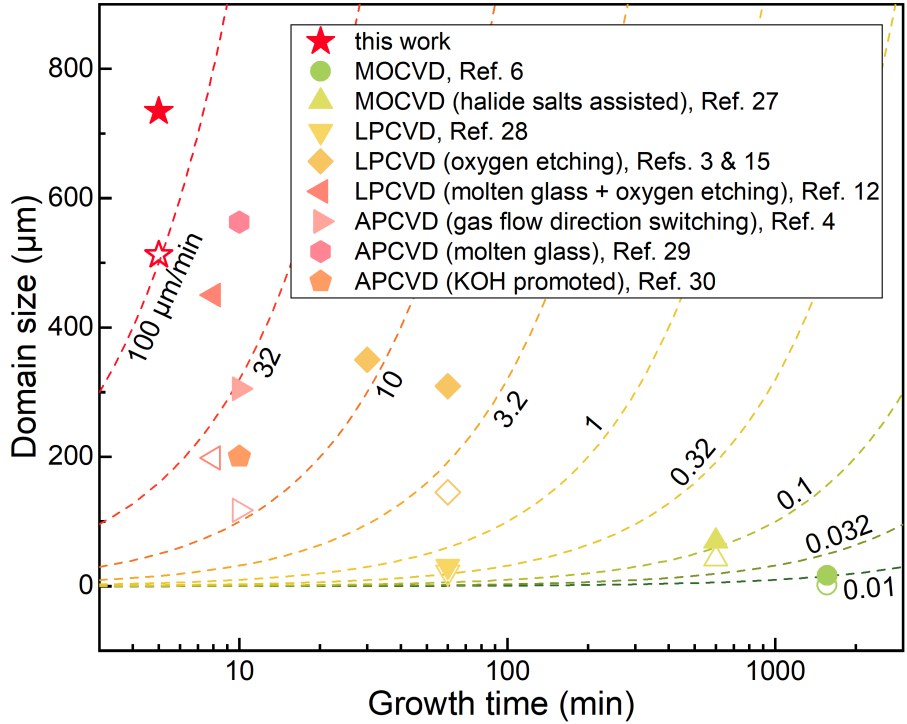

**Fig. 5: Benchmark of the optimized halide-assisted MoS₂ growth method.** The filled symbols describe the largest achievable domain sizes in these reports, while the corresponding hollow symbols mark the average domain sizes achieved. The "growth time" on $x$ axis indicates the time span of $T_{\text{Mo}}$. The dashed contour lines mark various growth velocities estimated by the domain size over the growth time. MOCVD: metal-organic CVD; LPCVD: low-pressure CVD; APCVD: ambient-pressure CVD.

**Ethics declarations**

The authors declare no competing financial interests.

**Acknowledgment**

Q.J. and J.K. acknowledge support by the STC Center for Integrated Quantum Materials, NSF Grant No. DMR-1231319. C.S. and J.K. acknowledge support the support from the U.S. Army Research Office (ARO) under grant no. W911NF-18-1-0431. N.M. and J.K. acknowledge support by U.S. Department of Energy, Office of Science, Basic Energy Sciences under Award DE-SC0020042. X.T. and J.M. acknowledge support by an Army Research Office MURI grant on Ab-Initio Solid-State Quantum Materials: Design, Production and Characterization at the Atomic Scale and by STROBE: an NSF Science and Technology Center under award DMR1548924. J.L. acknowledges support by an Office of Naval Research MURI through grant #N00014-17-1-2661. The work was also supported by the Director, Office of Science, Office of Basic Energy Sciences, Materials Sciences and Engineering Division, of the U.S. Department of Energy under Contract No. DE-AC02-05-CH11231, within the $sp^2$-Bonded Materials Program (KC2207), which supported supplementary modeling calculations. Additional support was provided by the National Science Foundation under Grant No. DMR-1807233 for TEM characterization. The STEM imaging is conducted at the Center for Nanophase Materials Sciences, which is a DOE Office of Science User Facility.